\documentclass[letterpaper, 10 pt, conference]{ieeeconf}  

\IEEEoverridecommandlockouts                              

\usepackage{color}
\usepackage{balance}

\usepackage{amsmath,amssymb,amsfonts}
\usepackage{algorithmic}
\usepackage{graphicx}
\usepackage{textcomp}
\usepackage{xcolor}
\usepackage{bm}
\usepackage{comment}
\usepackage{lineno,hyperref,float,amsfonts,amsmath,epsfig,graphics,booktabs,multirow,cite,stfloats}

\usepackage{amsthm,mathrsfs}

\theoremstyle{definition}

\theoremstyle{plain}

\theoremstyle{remark}

\def\BibTeX{{\rm B\kern-.05em{\sc i\kern-.025em b}\kern-.08em
    T\kern-.1667em\lower.7ex\hbox{E}\kern-.125emX}}
\AtBeginDocument{\definecolor{ojcolor}{cmyk}{0.93,0.59,0.15,0.02}}

\newcommand{\tabincell}[2]{\begin{tabular}{@{}#1@{}}#2\end{tabular}}

\title{\LARGE \bf Norwegian Electricity in Geographic Dataset (NoreGeo)

}
\author{Shiliang Zhang$^{1}$, Sabita Maharjan$^{1}$, Kai Strunz$^{2}$, Jan Christian Bryne$^{3}$
\thanks{This work was supported by the PriTEM project funded by UiO:Energy Convergence Environments.}
\thanks{We have openly released the generated dataset on Zenodo at \url{https://doi.org/10.5281/zenodo.16794603}, accessed Sep. 19, 2025.}
\thanks{$^{1}$Shiliang Zhang and Sabita Maharjan are with University of Oslo, Norway ({\tt\small \{shilianz, sabita\}@ifi.uio.no}).}
\thanks{$^{2}$Kar Strunz is with Technical University Berlin, Germany ({\tt\small kai.strunz@tu-berlin.de}).}
\thanks{$^{3}$Jan Christian Bryne is with Google Cloud Norway ({\tt\small bryne@google.com}).}
}

\begin{document}
\maketitle

\begin{abstract}

Geographic data is vital in understanding, analyzing, and contextualizing energy usage at the regional level within electricity systems. While geospatial visualizations of electricity infrastructure and distributions of production and consumption are available from governmental and third-party sources, these sources are often disparate, and compatible geographic datasets remain scarce. In this paper, we present a comprehensive geographic dataset representing the electricity system in Norway. We collect data from multiple authoritative sources, process it into widely accepted formats, and generate interactive maps based on this data. Our dataset includes information for each municipality in Norway for the year 2024, encompassing electricity infrastructure, consumption, renewable and conventional production, main power grid topology, relevant natural resources, and population demographics. This work results in a formatted geographic dataset that integrates diverse informational resources, along with openly released interactive maps. We anticipate that our dataset will alleviate software incompatibilities in data retrieval, and facilitate joint analyses on regional electricity system for energy researchers, stakeholders, and developers. \\

\end{abstract}

\section{BACKGROUND} 

The ongoing energy transition is driving the electricity system toward increased decarbonization, decentralization, and digitalization~\cite{10007664}. As nations endeavor to integrate variable distributed energy resources (DERs) and enhance energy efficiency~\cite{ADHAM20251980}, it becomes crucial to understand the intricate relationships between electricity infrastructure, resource availability, and demand patterns. Geospatial data analysis has emerged as a powerful tool for visualizing and examining these complex dynamics~\cite{SALVALAI2024114500,el2024comprehensive}, enabling a multi-layered representation of the various energy components involved~\cite{alhamwi2019development}. Valid geographic datasets in the electricity system is the key to facilitate informed decision-making in energy planning and scheduling, particularly under constraints related to resources and infrastructure~\cite{malinchik2010geo}. Quality open geospatial data is critical for ensuring the traceability and reproducibility of analyzed outcomes within the energy system~\cite{gotzens2019performing}.

Geospatial data for electricity systems can be accessed from authoritative or governmental sources, especially at the national level. However, these sources are often disparate and typically provide limited features, restricting utility of the data and hindering comprehensive analysis~\cite{sharma2023energy}. This motivates studies on retrieving, processing, and releasing electricity geographic datasets across various countries. The European Center for Medium-Range Weather Forecasts (ECMWF) released the ERA5 reanalysis dataset, which includes global resource data for wind and solar spanning the period 1979–2019~\cite{hersbach2019era5}. Weinand \textit{et al.} compiled 40 socio-energetic indicators across 11,131 municipalities in Germany to facilitate municipality level energy system assessments~\cite{weinand2019spatial}. Dunnett \textit{et al.} retrieved wind and solar power installations worldwide from OpenStreetMap and openly released the location data~\cite{dunnett2020harmonised}. Tavakkoli \textit{et al.} reviewed 146 spatiotemporal energy infrastructure datasets in USA, highlighting inter-agency collaboration and harmonization in data collection procedures~\cite{tavakkoli2021spatiotemporal}. Ortiz \textit{et al.} synthesized a location dataset for solar farms across India through satellite image analysis~\cite{ortiz2022artificial}. Chen \textit{et al.} estimated the energy consumption of 336 cities and 2,735 counties in China from multiple information sources for the period 1997-2017~\cite{chen2022city}. Jani \textit{et al.} assessed temporal and spatial simultaneity wind and solar energy resources in India~\cite{jani2022temporal} using the ERA5 reanalysis dataset. Sterl \textit{et al.} generated the dataset representing wind and solar power sites and their investment potential for the entire Africa continent~\cite{sterl2022all}. Antonio \textit{et al.} compiled the dataset documenting installed renewable energy in Spain from 2015 to 2020, covering 59,386 solar PV plants and 1,205 wind farms~\cite{jimenez2023sowisp}. Maneesh \textit{et al.} integrated datasets related to carbon capture and storage (CCS) in USA from various federal agencies, encompassing social and environmental factors that can influence the viability of CCS projects~\cite{sharma2023energy}. Yang \textit{et al.} estimated the energy consumption of 331 cities in China during the period 2005-2021~\cite{yang2024comprehensive}. Moya \textit{et al.} synthesized an energy consumer dataset incorporating geospatial, socioeconomic, and demographic factors~\cite{moya2024global}. The aforementioned collected or synthesized data can enhance geospatial analysis and support realistic decision-making in the energy system. Nevertheless, there remains a scarcity of compatible datasets that can be readily processed in Geographic Information Systems (GIS) platforms, such as QGIS or ArcGIS. The compatibility of dataset is crucial since it releases nontrivial efforts in reformatting the data to fit targeting GIS platforms, thus facilitating direct usage and further customization of the data for individual analyses. Furthermore, there is a lack of openly accessible interactive maps based on geographic energy datasets, which creates technical barriers that impede intuitive understanding and reasoning of the data among energy stakeholders.     

In this paper, we provide an openly accessible geographic dataset that characterizes the electricity system in Norway for the year 2024, accompanied by interactive maps derived from the dataset. We compile data relevant to the Norwegian electricity system and energy usage from authoritative sources to capture a diverse array of geographic and demographic features. The data sources utilized include Statistics Norway (SSB)\footnote{\url{https://www.ssb.no/en}, accessed Sep. 19, 2025.}, the national statistical institute; Geonorge\footnote{\url{https://www.geonorge.no/en/}, accessed Sep. 19, 2025.}, the national platform for map data; the NVE Kartkatalog\footnote{\url{https://kartkatalog.nve.no}, accessed Sep. 19, 2025.}, the map catalog managed by the Norwegian Water Resources and Energy Directorate; eSett\footnote{\url{https://www.esett.com}, accessed Sep. 19, 2025.}, which offers imbalance settlement services to electricity market participants in Nordic countries; and OpenStreetMap\footnote{\url{https://www.openstreetmap.org}, accessed Sep. 19, 2025.}. We process the collected data into CSV\footnote{Comma-separated values (CSV) is a plain text file format to store tabular data, such as spreadsheets or databases} and GeoJSON\footnote{GeoJSON is an open standard format designed for representing geographical features.} files compatible with GIS platforms through the use of QGIS~\cite{moyroud2018introduction}, ArcGIS~\cite{johnston2001using}, Python, Google Colab~\cite{bisong2019google}, and Overpass turbo\footnote{\url{https://overpass-turbo.eu}, accessed Sep. 19, 2025.}. The features of our released dataset include (i) power production, including power generation from hydro, wind, and solar, and wind and hydro power resources, (ii) power grid and infrastructure, including the topology and capacity of main power lines, transformers and their capacities, hydro, wind, and solar power plants and their specifications, hydro power pipes and tunnels, distribution of solar panels, and (iii) power consumption, including municipality level energy consumption, and population distribution that is closely related to the consumption. Additionally, we generate interactive maps based on our dataset, allowing users to access and customize the maps for specific use cases. We anticipate this dataset and the interactive maps to serve as a valuable resource for researchers and stakeholders engaged in joint analysis of electricity system and energy policy at national levels, \textit{e.g.}, in the applications of infrastructure and capacity planning, grid vulnerability prognosis, electricity transmission scheduling, and risk monitoring in power supply, \textit{etc}. While existing datasets can contribute to those applications, this paper provides a formatted and compatible dataset that facilitates direct use along with intuitive understanding of the data.

\section{COLLECTION METHODS AND DESIGN} 

\begin{figure*}[htb]
\centerline{\includegraphics[width=6.3in]{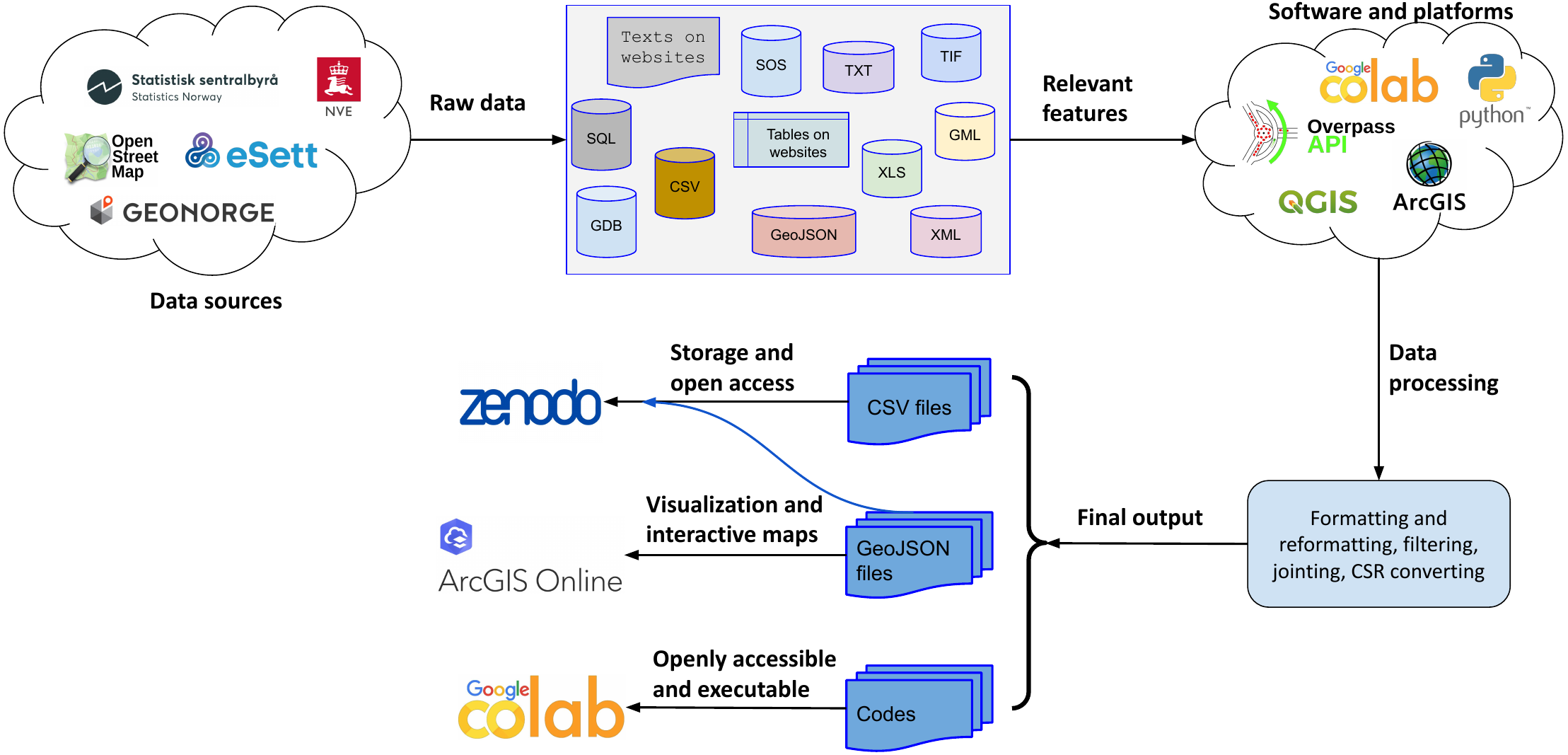}}
\caption{Diagram for the whole procedure of data collection and processing.\label{fig1}}
\end{figure*}

We collect the raw data from multiple authoritative open source resources in different formats. From the raw data, we select features relevant to the energy usage and infrastructure of the Norwegian electricity system. We then reformat and process the data using several software and platforms to produce the final outputs in CSV and GeoJSON formats. We use the generated GeoJSON data to build interactive maps. We provide all the data, the code used to processing the data, and the interactive maps as openly accessible resources. We show the overall data collection procedures in Fig.~\ref{fig1}, and we explain details in generating the data in the following.

\subsection{Energy consumption data}

\begin{figure}[htb]
\centerline{\includegraphics[width=3.3in]{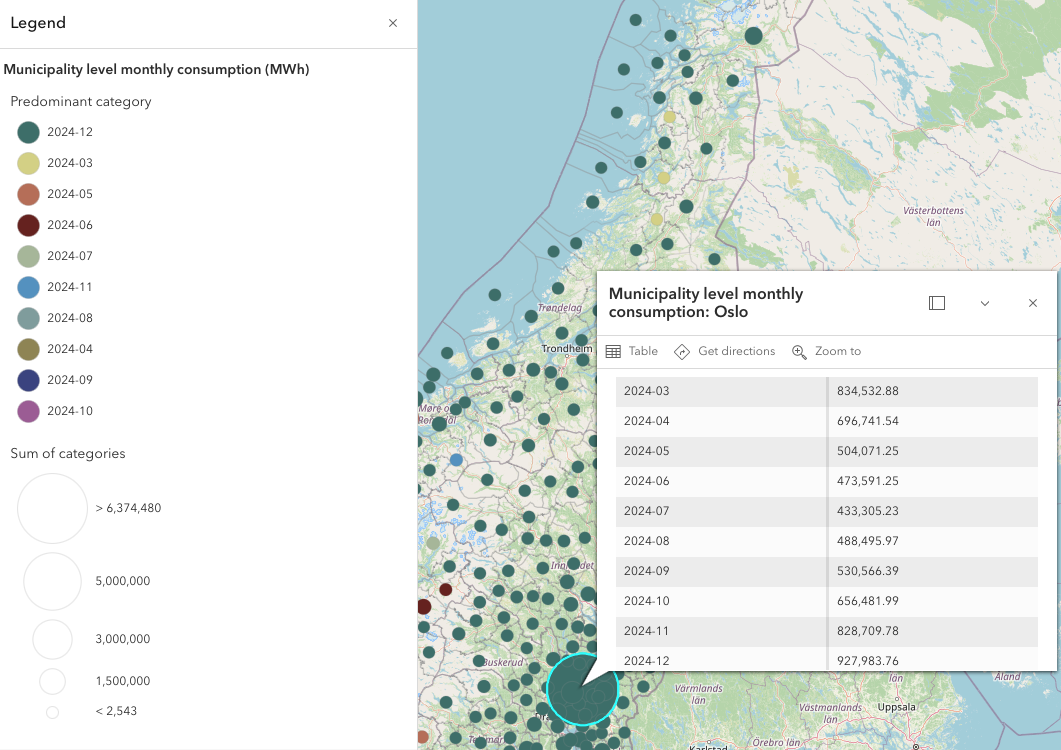}}
\caption{Norwegian monthly municipality-level energy consumption (MWh) in 2024.\label{fig:energy_consumption}}
\end{figure}

We collected the raw data from NVE in XLSX file for municipality level power consumption in Norway\footnote{The raw data is publicly available at \url{https://www.nve.no/energi/energisystem/energibruk/energibruk-i-kommuner/}, accessed Sep. 19, 2025.}. The raw data contains name of the 357 Norwegian municipalities and their monthly power consumption in 2024. We only retrieved the consumption data for one year to keep it consistent with the period where we collect the other features. Note that for the monthly consumption, the data is only available from March to December.

We combined this data with the geographic boundary of Norwegian municipalities accessed from Geonorge.no\footnote{The raw data is publicly available at \url{https://kartkatalog.geonorge.no/metadata/administrative-units-municipalities/041f1e6e-bdbc-4091-b48f-8a5990f3cc5b}, accessed Sep. 19, 2025.}. We retrieved the raw data in GeoJSON file, and we generated Python code in Google Colab to integrate the geographic boundaries with the power consumption for each Norwegian municipality. The final output of the municipality level energy consumption data is available in both CSV and GeoJSON format at Zenodo. The interactive map (see the screenshot in Fig.~\ref{fig:energy_consumption}) we built based on this data is openly released via Online ArchGIS\footnote{The interactive map is publicly available at \url{https://uio-no.maps.arcgis.com/apps/mapviewer/index.html?webmap=f1efdace78c741599b34271749afaf14}, accessed Sep. 19, 2025.}.

\subsection{Electricity price}

\begin{figure}[htb]
\centerline{\includegraphics[width=3.3in]{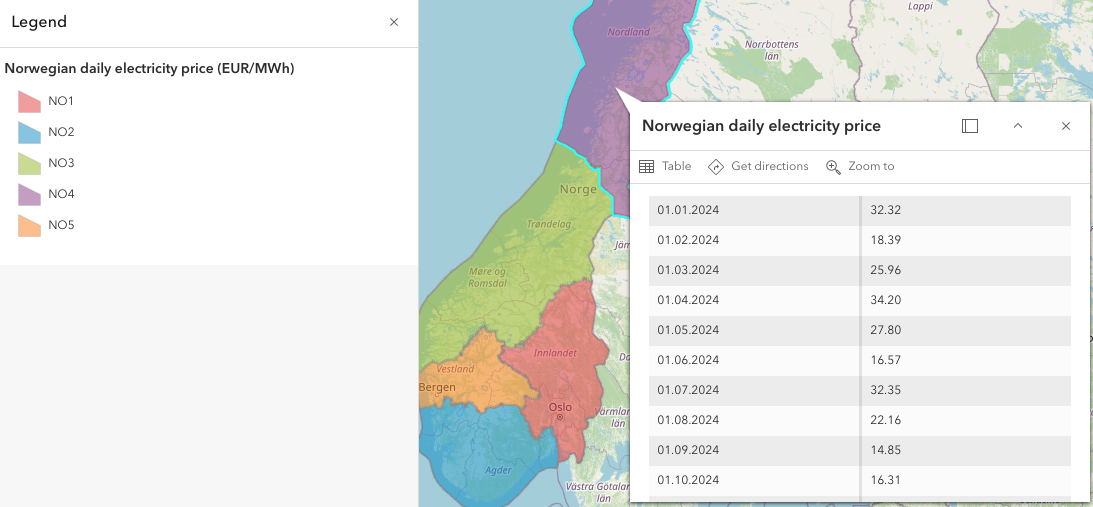}}
\caption{Electricity price (EUR/MWh) in 2024 in Norway. Note that the time is in the format of DD-MM-YYYY.\label{fig:price}}
\end{figure}

There are five market balance areas (MBAs) in Norway, each of which has its own electricity price. We collected the geographic boundaries of the five MBA areas, and combined them with their electricity prices (in EUR/MWh) in 2024 on a daily basis. We retrieved the MBA boundaries from NVE Kartkatalog\footnote{The raw data is publicly available at \url{https://temakart.nve.no/tema/nettanlegg}, accessed Sep. 19, 2025.} in GeoJSON format\footnote{SHAPE and KMZ formats are also available for this data.}, and we obtained electricity price data from eSett\footnote{The raw data is publicly availabe at \url{https://opendata.esett.com/prices_single}, accessed Sep. 19, 2025.} in CSV format. We integrated the MBA boundary and the price data using Python, and generated the dataset in CSV and GeoJSON data that present the daily electricity price in Norway during 2024, as shown in Fig.~\ref{fig:price}. The final output of the dataset is released at Zenodo, and we visualize the MBAs and their daily electricity prices by an interactive map via Online ArcGIS\footnote{The interactive map is publicly accessible at \url{https://uio-no.maps.arcgis.com/apps/mapviewer/index.html?webmap=6c80f11c8411479d8b518f86d90c0048}, accessed Sep. 19, 2025.}. 

\subsection{Population density}

\begin{figure}[tb]
\centerline{\includegraphics[width=3.3in]{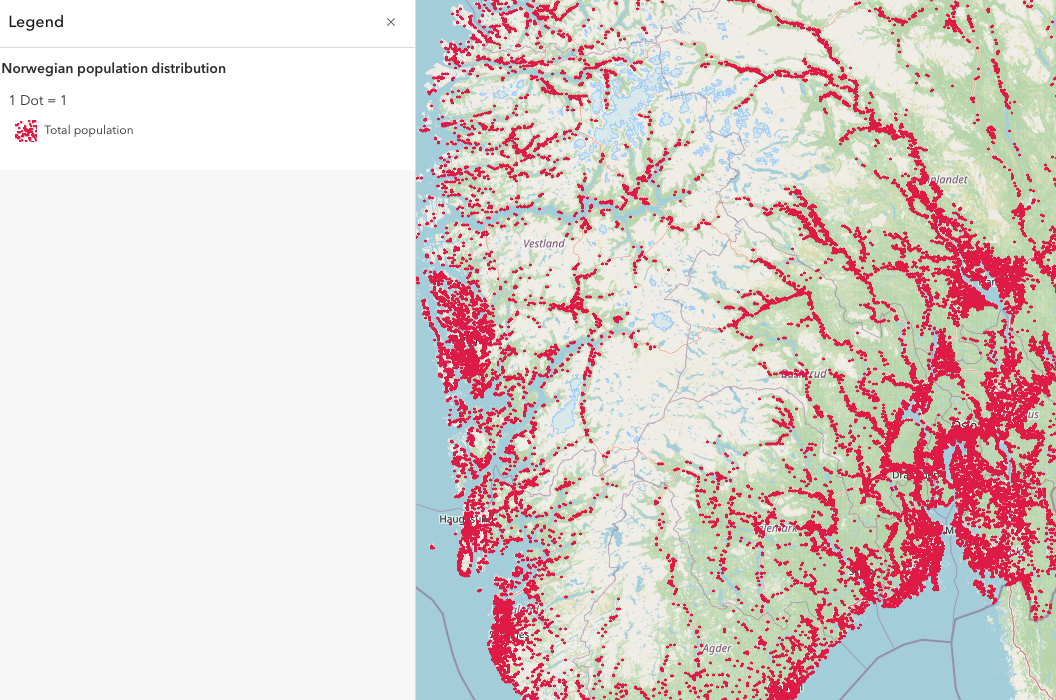}}
\caption{Norwegian population density in 250 meter resolution in 2024.\label{fig:population_density}}
\end{figure}

We retrieved the population density of Norway from Geonorge.no\footnote{The raw data is publicly available at \url{https://kartkatalog.geonorge.no/metadata/befolkning-paa-rutenett-250-m-2024/e668abf5-2fac-472e-976b-1526f1d38bbe}, accessed Sep. 19, 2025.} which is based on the data from SSB. We consider demographic data in this paper since the number of residents in an area significantly influences the consumption, particularly during the winter when heating in residential buildings is one of the major sources of energy consumption. This data presents the population density of the whole Norway with the resolution 250m $\times$ 250m for the year 2024. The raw data is in GML format, and we used QGIS to convert it into GeoJSON and further process it to CSV file using Python. A visualization of the generated GeoJSON data is shown in Fig.~\ref{fig:population_density}. We created interactive map for this data via Online ArcGIS\footnote{The map is accessible at \url{https://uio-no.maps.arcgis.com/apps/mapviewer/index.html?webmap=3a5917deab76451e944936d75404234c}, accessed Sep. 19, 2025.}.

\subsection{Main power grid}

\begin{figure}[tbp]
\centerline{\includegraphics[width=3.3in]{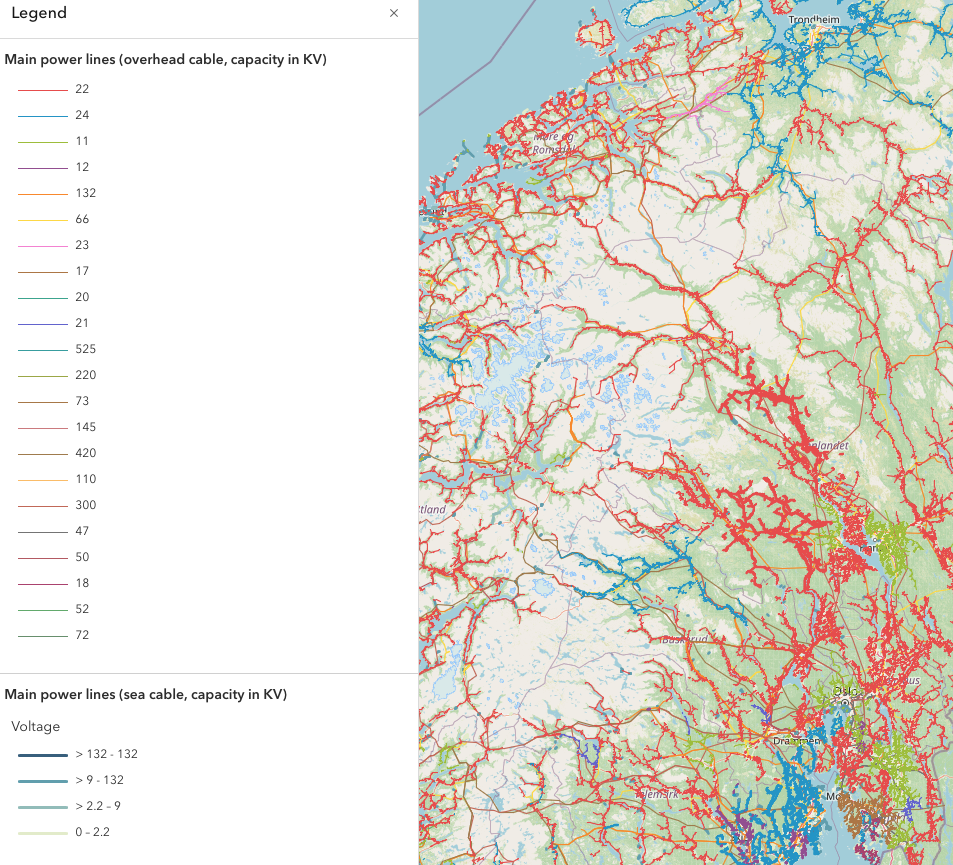}}
\caption{Norwegian main power grid topology and capacity.\label{fig:power_grid}}
\end{figure}

\begin{figure}[tbp]
\centerline{\includegraphics[width=3.3in]{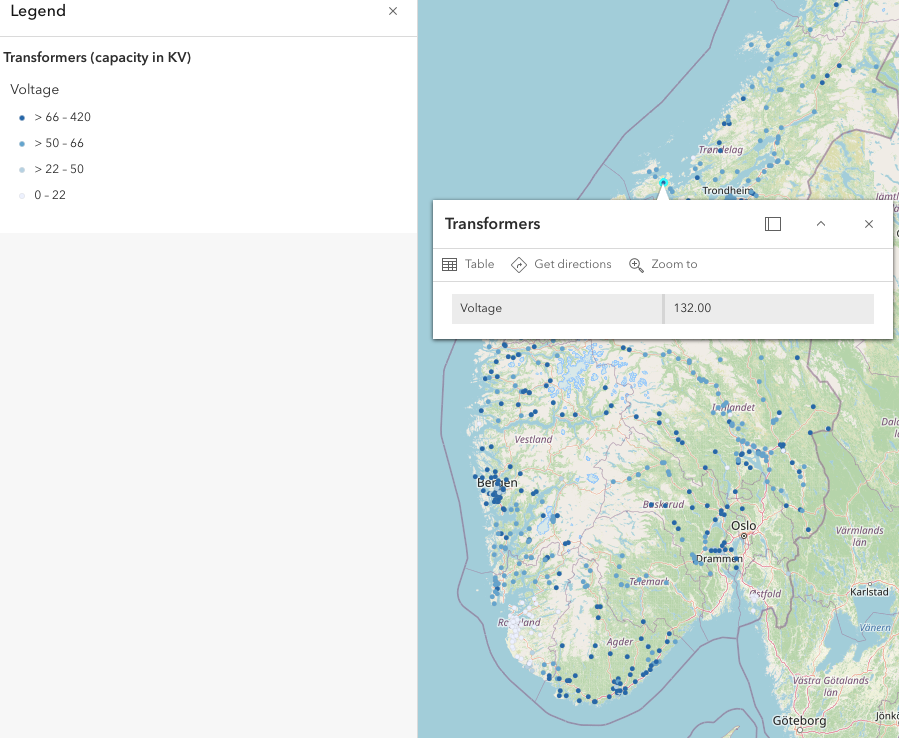}}
\caption{Transformers and their capacities in the Norwegian power grid.\label{fig:transformers}}
\end{figure}

We collected the main power grid data across Norway from Geonorge.no\footnote{The raw data is publicly avialable at \url{https://kartkatalog.geonorge.no/metadata/nettanlegg/9f71a24b-9997-409f-8e42-ce6f0c62e073}, accessed Sep. 19, 2025.} in GML format\footnote{This data is also available in SOS, SQL, and GDB format.}. The data contains the topology and location of overhead cables (32-525kV), submarine cables (32-170kV), and transformer stations (24-525kV) in transmission networks, regional and high-voltage distribution networks. The power lines and transformers are associated with their capacity in the data. We converted the collected GML data into GeoJSON format using QGIS. The final output is stored in Zenodo and visualized in interactive maps via Online ArcGIS\footnote{The interactive maps for main power grid and transformers are available at \url{https://uio-no.maps.arcgis.com/apps/mapviewer/index.html?webmap=8e4906551c3346249cfd443ca70507cd} and \url{https://uio-no.maps.arcgis.com/apps/mapviewer/index.html?webmap=71ad265a41804be3a66dff39d24d0548}, accessed Sep. 19, 2025.}, as shown in Fig.~\ref{fig:power_grid} (main power grid) and Fig.~\ref{fig:transformers} (transformers).

\subsection{Hydro power, power resources, pipes, and tunnels}

\begin{figure}[tbp]
\centerline{\includegraphics[width=3.3in]{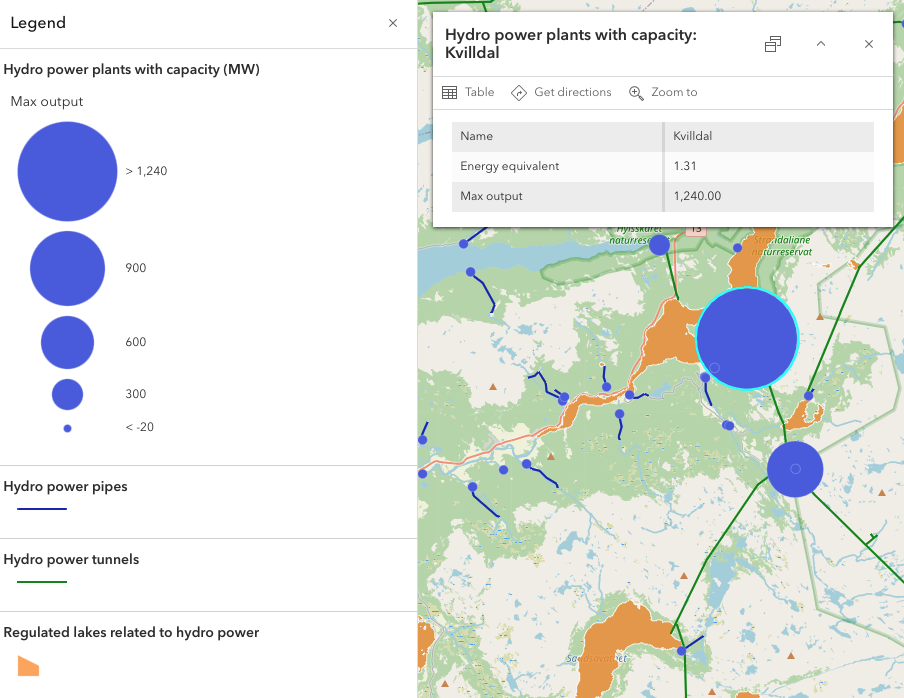}}
\caption{Hydro power plants and capacities, regulated lakes related to hydro resources, and hydro power pipes and tunnels.\label{fig:hydro}}
\end{figure}

We retrieved the raw data from Geonorge.no\footnote{The raw data is publicly available at \url{https://kartkatalog.geonorge.no/metadata/vannkraft-utbygd-og-ikke-utbygd/f587a15a-c72a-4b21-aae9-4132df1bdd27}, accessed Sep. 19, 2025.} in GML format\footnote{This data is also available in GDB, SQL, and SOS format.}. We selected the features including the hydro power plant capacity (in MW), regulated lakes that affect the watercourses, and pipes and tunnels facilitating the hydro power production. We also included lengths of the pipes and tunnels. We considered hydro power plants both in operation and not. Note that there may be power plants not in operation due to license application process. We converted the GML data into GeoJSON format using QGIS, and we created an interactive map for this data using ArcGIS, as shown in Fig.~\ref{fig:hydro}. The interactive map is openly accessible via Online ArcGIS\footnote{The interactive map is available at \url{https://uio-no.maps.arcgis.com/apps/mapviewer/index.html?webmap=b636fc5747ec4e7f8c02f1c9af582990}, accessed Sep. 19, 2025.}.

\subsection{Solar power production, power plants, planned solar power plan sites, and city-level PV installation}

\begin{figure}[htb]
\centerline{\includegraphics[width=3.3in]{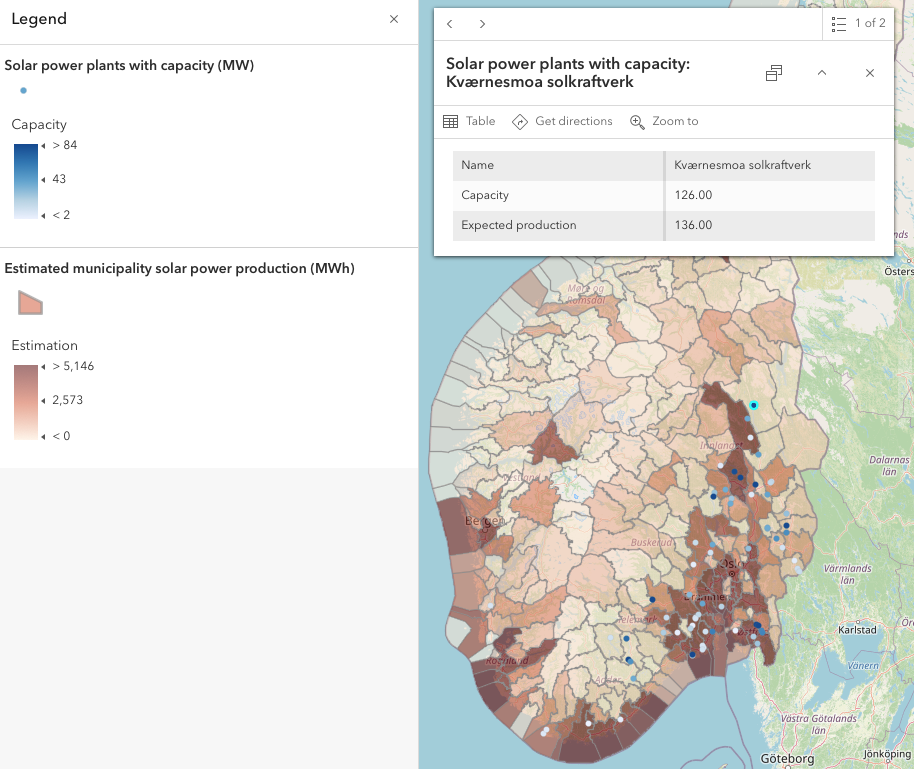}}
\caption{Municipality level solar power production estimation and power plants with capacity.\label{fig:solar}}
\end{figure}

We collected the estimated yearly municipality-level solar power production from NVE\footnote{The raw data is publicly avaiable at \url{https://www.nve.no/energi/energisystem/solkraft/oversikt-over-solkraftanlegg-i-norge/}, accessed Sep. 19, 2025.} in XLSX format. Note that the actual solar power production is unknown, since the power generation from individual rooftops is not metered. The production data provides a simplified representation of the solar power production in an average weather year by NVE. We converted this data into GeoJSON format using Python and combined it with the municipal geographic boundaries to generated the data readable in GIS platforms. 

Furthermore, we collected Norwegian solar power plants and their capacity data from NVE\footnote{The raw data is publicly available at \url{https://temakart.nve.no/tema/solkraftverk}, accessed Sep. 19, 2025.} in GeoJSON format\footnote{The raw data is also available in CSV, KMZ, JSON, and SHAPE formats.}. The data includes solar power plants that are subject to completed license or with license application under processing in Norway. We visualized the solar power plants and estimated municipality-level solar power production in an interactive map via Online ArcGIS\footnote{The interactive map is publicly accessible at \url{https://uio-no.maps.arcgis.com/apps/mapviewer/index.html?webmap=469df60bcb374f0b997cf7fd516fc3a6}, accessed Sep. 19, 2025.}, and we show how it looks like in Fig.~\ref{fig:solar}.

\begin{figure}[htb]
\centerline{\includegraphics[width=3.3in]{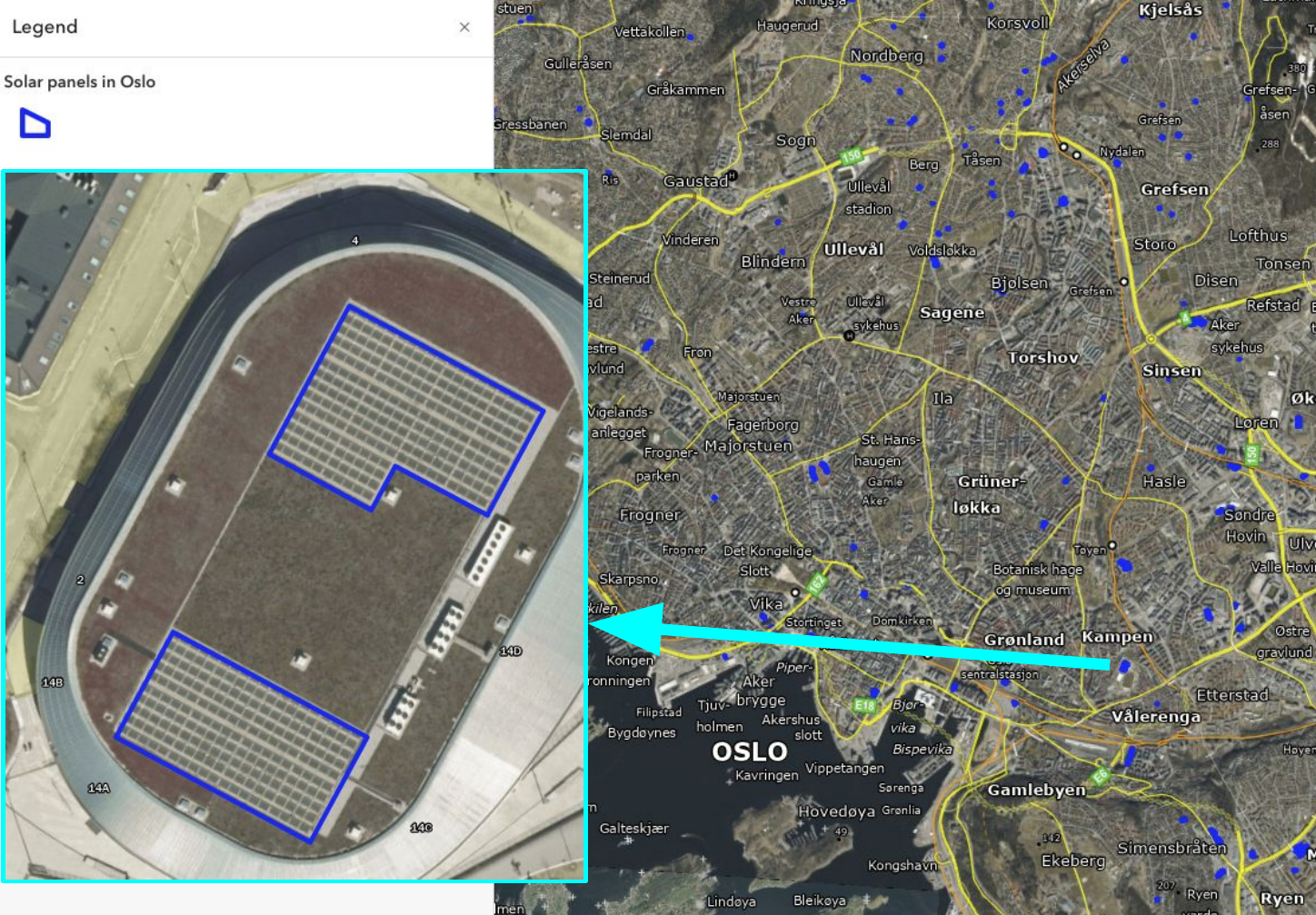}}
\caption{Distribution of solar panels in the city of Oslo. Solar panels are outlined in blue.\label{fig:solar_pv}}
\end{figure}

We also provide an example of how solar panels are distributed in a city. This distribution can be used to check the geographic context of solar panel installation, particularly for residential rooftops. We retrieved the data from Open Street Map\footnote{\url{https://www.openstreetmap.org/}, accessed Sep. 19, 2025.} using Python and the Overpass turbo API\footnote{\url{https://overpass-turbo.eu}, accessed Sep. 19, 2025.}. The data includes polygons indicating areas covered by rooftop solar panels in the city of Oslo in Norway. The solar panels accounts an area of 104,024.40 square meters in total in Oslo.

We only collected the data for Oslo, as shown in Fig.~\ref{fig:solar_pv}, while the approach for the data retrieval and processing can be applicable for any other city in the world. We stored the processed data at Zenodo, and we visualized the data by an interactive map via Online ArcGIS\footnote{The interactive map is publicly accessible at \url{https://uio-no.maps.arcgis.com/apps/mapviewer/index.html?webmap=045e64206c874c0eb5bb5e1e2e7cd619}, accessed Sep. 19, 2025.}.

\subsection{Wind power plant data and wind resource availability}

\begin{figure}[htb]
\centerline{\includegraphics[width=3.3in]{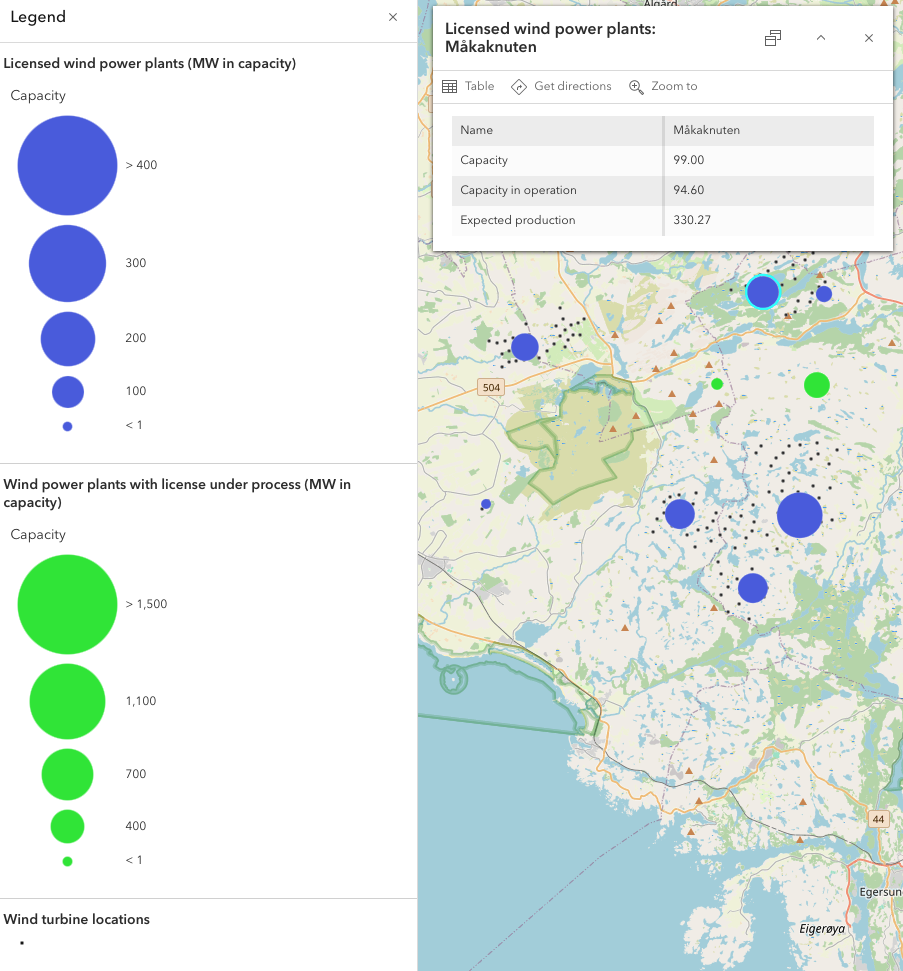}}
\caption{wind power plants and turbines in Norway.\label{fig:wind_power_plants}}
\end{figure}

\begin{figure}[htb]
\centerline{\includegraphics[width=3.3in]{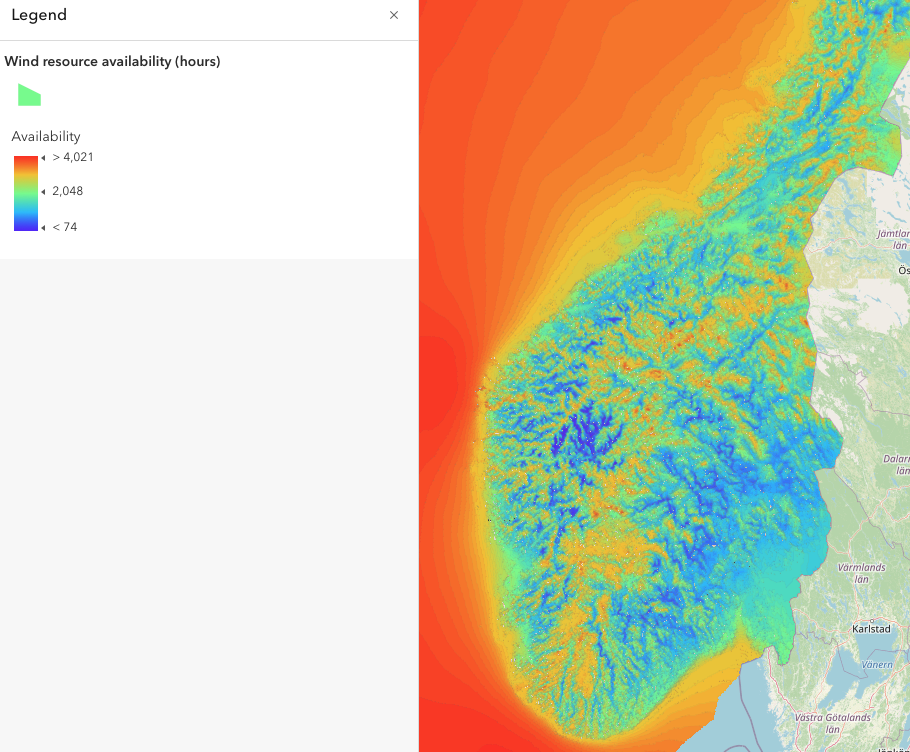}}
\caption{Wind resources measured by turbine operating hours in a year at the height of 50 meters above ground.\label{fig:wind_resource}}
\end{figure}

\begin{table}[tbhp]%
	\centering \caption{\label{tbl:data_validity}Information regarding validity of our dataset.}
	\resizebox{0.48\textwidth}{!}{
		\begin{tabular}{|l|l|l|l|l|}
			\cline{1-5}
			\tabincell{l}{\textbf{Name of the data}} & \tabincell{l}{\textbf{Data source}} & \tabincell{l}{\textbf{Date of}\\ \textbf{creation}} & \tabincell{l}{\textbf{Data type}\\ \textbf{in the source}} & \tabincell{l}{\textbf{Accuracy}} \\

            \cline{1-5}
			\tabincell{l}{Electricity price} & \tabincell{l}{eSett} & \tabincell{l}{Near\\ real-time} & \tabincell{l}{Real and public} & \tabincell{l}{Accurate} \\
			\cline{1-5}
			\tabincell{l}{Power consumption}& \tabincell{l}{NVE} & \tabincell{l}{01.04.2025} & \tabincell{l}{Real and registered} & \tabincell{l}{Accurate} \\
			\cline{1-5}
			\tabincell{l}{Power grids and \\transformers} & \tabincell{l}{Geonorge.no} & \tabincell{l}{06.05.2025} & \tabincell{l}{Real and registered} & \tabincell{l}{Accurate} \\
            \cline{1-5}
			\tabincell{l}{Municipal boundary} & \tabincell{l}{Geonorge.no} & \tabincell{l}{17.12.2014} & \tabincell{l}{Real and public} & \tabincell{l}{Accurate} \\
			\cline{1-5}
			\tabincell{l}{Population density} & \tabincell{l}{SSB and \\Geonorge.no} & \tabincell{l}{11.12.2024} & \tabincell{l}{Sampled and estimated} & \tabincell{l}{High} \\
            \cline{1-5}
			\tabincell{l}{Electricity price \\areas} & \tabincell{l}{NVE} & \tabincell{l}{01.04.2025} & \tabincell{l}{Real and public} & \tabincell{l}{Accurate} \\
            \cline{1-5}
			\tabincell{l}{Hydro power plants\\Hydro pipes and \\tunnels, Regulated \\lakes} & \tabincell{l}{Geonorge.no} & \tabincell{l}{06.05.2025} & \tabincell{l}{Real and registered} & \tabincell{l}{Accurate} \\
            \cline{1-5}
			\tabincell{l}{Wind power plants \\and turbines} & \tabincell{l}{Geonorge.no} & \tabincell{l}{20.08.2025} & \tabincell{l}{Real and registered} & \tabincell{l}{Accurate} \\
            \cline{1-5}
			\tabincell{l}{Wind resource \\availability} & \tabincell{l}{NVE and \\Geonorge.no} & \tabincell{l}{15.11.2023} & \tabincell{l}{Sampled and estimated} & \tabincell{l}{High} \\
            \cline{1-5}
			\tabincell{l}{Solar power plants} & \tabincell{l}{NVE} & \tabincell{l}{06.05.2025} & \tabincell{l}{Real and registered} & \tabincell{l}{Accurate} \\
            \cline{1-5}
			\tabincell{l}{Municipality solar \\power production} & \tabincell{l}{NVE} & \tabincell{l}{31.12.2024} & \tabincell{l}{Estimated} & \tabincell{l}{Moderate} \\
            \cline{1-5}
			\tabincell{l}{Solar panel \\distribution in Oslo} & \tabincell{l}{Open Street \\Map} & \tabincell{l}{02.05.2025} & \tabincell{l}{Individual observation} & \tabincell{l}{Moderate} \\
            \cline{1-5}
            \multicolumn{5}{|l|}{\tabincell{l}{\textbf{Real and public} indicates real data which is made transparent and accessible to the public\\ by governmental institutes.}}\\
            \multicolumn{5}{|l|}{\tabincell{l}{\textbf{Real and registered} means real data that is registered and reported to government by\\ energy stakeholders.}}\\
            \multicolumn{5}{|l|}{\tabincell{l}{\textbf{Sampled and estimated} indicates estimated data through sampling and statistical measures.}}\\
            \multicolumn{5}{|l|}{\tabincell{l}{\textbf{Estimated} indicates estimation data according to reasonable assumptions and conditions.}}\\
            \multicolumn{5}{|l|}{\tabincell{l}{\textbf{Individual observations} means data contributed by individuals in open-source communities.}}\\
            \cline{1-5}

		\end{tabular}
	}
\end{table}

\begin{table*}[tbhp]%
	\centering \caption{\label{tbl:csv_details}Summary of details in the generated dataset files.}
	\resizebox{0.99\textwidth}{!}{
		\begin{tabular}{|l|l|l|l|l|}
			\cline{1-5}
			\tabincell{l}{File name} & \tabincell{l}{Number \\of records} & \tabincell{l}{Feature name} & \tabincell{l}{Meaning} & \tabincell{l}{Unit}  \\

            \cline{1-5}
			\multirow{3}{*}{\tabincell{l}{Norwegian daily electricity price.csv}} & \multirow{3}{*}{\tabincell{l}{1,830}} & \tabincell{l}{Date CET/CEST} & \tabincell{l}{Time according to CET/CEST} & \tabincell{l}{NA} \\
			\cline{3-5}
			&  & \tabincell{l}{MBA} & \tabincell{l}{Market balance area, or the electricity price area in Norway} & \tabincell{l}{NA} \\
			\cline{3-5}
			&  & \tabincell{l}{spot\_price} & \tabincell{l}{Electricity price} & \tabincell{l}{Eur per MWh} \\

            \cline{1-5}
			\multirow{3}{*}{\tabincell{l}{Norwegian daily electricity price.geojson}} & \multirow{3}{*}{\tabincell{l}{1,830}} & \tabincell{l}{Electricity\_price\_area} & \tabincell{l}{Market balance area, or the electricity price area in Norway} & \tabincell{l}{NA} \\
			\cline{3-5}
			&  & \tabincell{l}{Shape\_length} & \tabincell{l}{Length of the boundary of the electricity price area} & \tabincell{l}{Meter} \\
			\cline{3-5}
			&  & \tabincell{l}{Shape\_area} & \tabincell{l}{Area of the electricity price area} & \tabincell{l}{Square meter} \\
            \cline{3-5}
            &  & \tabincell{l}{Electricity price and time stamp} & \tabincell{l}{Time in the format DD.MM.YYYY} & \tabincell{l}{Electricity meansured by\\ Eur/MWh} \\
            \cline{3-5}
            &  & \tabincell{l}{Geometry} & \tabincell{l}{Polygon of the geographic boundary of the electricity price area with latitude and longitude} & \tabincell{l}{NA} \\

            \cline{1-5}
			\multirow{3}{*}{\tabincell{l}{Municipality level monthly consumption.csv}} & \multirow{3}{*}{\tabincell{l}{3,580}} & \tabincell{l}{Month} & \tabincell{l}{Month during the year 2024} & \tabincell{l}{NA} \\
			\cline{3-5}
			&  & \tabincell{l}{Municipality} & \tabincell{l}{Name of the municipalities in Norway} & \tabincell{l}{NA} \\
			\cline{3-5}
			&  & \tabincell{l}{Total energy consumption} & \tabincell{l}{The amount of energy consumed within a month} & \tabincell{l}{MWh} \\

            \cline{1-5}
			\multirow{3}{*}{\tabincell{l}{Municipality level monthly consumption.geojson}} & \multirow{3}{*}{\tabincell{l}{3,580}} & \tabincell{l}{Municipality name} & \tabincell{l}{Name of municipalities in Norway} & \tabincell{l}{NA} \\
			\cline{3-5}
			&  & \tabincell{l}{Energy consumption and time stamp} & \tabincell{l}{Time in the format YYYY-MM} & \tabincell{l}{Consumption measured \\by MWh} \\
			\cline{3-5}
			&  & \tabincell{l}{Geometry} & \tabincell{l}{Polygon of the geographic boundary of municipalities in Norway with latitude and longitude} & \tabincell{l}{NA} \\

            \cline{1-5}
			\multirow{2}{*}{\tabincell{l}{Main power lines (overhead cable).geojson}} & \multirow{2}{*}{\tabincell{l}{145,891}} & \tabincell{l}{Voltage} & \tabincell{l}{Voltage of the electricity through the cable} & \tabincell{l}{KV} \\
			\cline{3-5}
			&  & \tabincell{l}{Geometry} & \tabincell{l}{The power line on the map with latitudes and longitudes} & \tabincell{l}{NA} \\

            \cline{1-5}
			\multirow{2}{*}{\tabincell{l}{Main power lines (sea cable).geojson}} & \multirow{2}{*}{\tabincell{l}{8,762}} & \tabincell{l}{Voltage} & \tabincell{l}{Voltage of the electricity through the cable} & \tabincell{l}{KV} \\
			\cline{3-5}
			&  & \tabincell{l}{Geometry} & \tabincell{l}{The power line on the map with latitudes and longitudes} & \tabincell{l}{NA} \\

            \cline{1-5}
			\multirow{2}{*}{\tabincell{l}{Transformers.geojson}} & \multirow{2}{*}{\tabincell{l}{1,211}} & \tabincell{l}{Voltage} & \tabincell{l}{Voltage of the electricity through the transformer} & \tabincell{l}{KV} \\
			\cline{3-5}
			&  & \tabincell{l}{Geometry} & \tabincell{l}{The transformer location on the map in latitude and longitude} & \tabincell{l}{NA} \\

            \cline{1-5}
			\multirow{2}{*}{\tabincell{l}{Norwegian municipality boundaries.geojson}} & \multirow{2}{*}{\tabincell{l}{357}} & \tabincell{l}{Municipality name} & \tabincell{l}{Name of the municipalities in Norway} & \tabincell{l}{NA} \\
			\cline{3-5}
			&  & \tabincell{l}{Geometry} & \tabincell{l}{Polygon of the municipality boundary in latitudes and longitudes} & \tabincell{l}{NA} \\

            \cline{1-5}
			\multirow{1}{*}{\tabincell{l}{Norwegian market balance area boundaries.geojson}} & \multirow{1}{*}{\tabincell{l}{5}} & \tabincell{l}{Geometry} & \tabincell{l}{Polygon of the geographic boundary of the electricity price area with latitudes and longitudes} & \tabincell{l}{NA} \\

            \cline{1-5}
			\multirow{2}{*}{\tabincell{l}{Norwegian population distribution.geojson}} & \multirow{2}{*}{\tabincell{l}{224,541}} & \tabincell{l}{Total population} & \tabincell{l}{Number of people living in a grid of 250m$\times$250m} & \tabincell{l}{NA} \\
			\cline{3-5}
			&  & \tabincell{l}{Geometry} & \tabincell{l}{The latitudes and longitudes for the geographic boundary of a grid with the size 250m$\times$250m} & \tabincell{l}{NA} \\

            \cline{1-5}
			\multirow{3}{*}{\tabincell{l}{Hydro power plants with capacity.geojson}} & \multirow{3}{*}{\tabincell{l}{4,052}} & \tabincell{l}{Max output} & \tabincell{l}{Maximum output of the power plant} & \tabincell{l}{MW} \\
			\cline{3-5}
			&  & \tabincell{l}{Name} & \tabincell{l}{Name of the hydro power plant} & \tabincell{l}{NA} \\
			\cline{3-5}
			&  & \tabincell{l}{Energy equivalent} & \tabincell{l}{It means how much energy can be extracted per cubic meter of water through the power plant} & \tabincell{l}{KWh/m$^3$} \\
            \cline{3-5}
			&  & \tabincell{l}{Geometry} & \tabincell{l}{Location of the power plant} & \tabincell{l}{NA} \\

            \cline{1-5}
			\multirow{2}{*}{\tabincell{l}{Hydro power pipes.geojson}} & \multirow{2}{*}{\tabincell{l}{3,678}} & \tabincell{l}{Length} & \tabincell{l}{The length of the pipe} & \tabincell{l}{Meter} \\
			\cline{3-5}
			&  & \tabincell{l}{Geometry} & \tabincell{l}{The pipes on the map with latitutes and longitudes} & \tabincell{l}{NA} \\

            \cline{1-5}
			\multirow{2}{*}{\tabincell{l}{Hydro power tunnels.geojson}} & \multirow{2}{*}{\tabincell{l}{2,839}} & \tabincell{l}{Length} & \tabincell{l}{Length of the hydro power plant tunnels} & \tabincell{l}{Meter} \\
			\cline{3-5}
			&  & \tabincell{l}{Geometry} & \tabincell{l}{The tunnels on the map with latitudes and longitudes} & \tabincell{l}{Meter} \\

            \cline{1-5}
			\multirow{2}{*}{\tabincell{l}{Regulated lakes related to hydro power.geojson}} & \multirow{2}{*}{\tabincell{l}{2,448}} & \tabincell{l}{Name} & \tabincell{l}{Name of the regulated lake} & \tabincell{l}{NA} \\
			\cline{3-5}
			&  & \tabincell{l}{Geometry} & \tabincell{l}{Polygon of the lake with latitudes and longitudes} & \tabincell{l}{NA} \\

            \cline{1-5}
			\multirow{3}{*}{\tabincell{l}{Licensed wind power plants.geojson}} & \multirow{3}{*}{\tabincell{l}{64}} & \tabincell{l}{Capacity} & \tabincell{l}{Power capacity of the plant when designed} & \tabincell{l}{MW} \\
            \cline{3-5}
			&  & \tabincell{l}{Name} & \tabincell{l}{Name of the power plant} & \tabincell{l}{NA} \\
			\cline{3-5}
			&  & \tabincell{l}{Capacity in operation} & \tabincell{l}{Capacity of the power plant during operation} & \tabincell{l}{MW} \\
			\cline{3-5}
			&  & \tabincell{l}{Expected production} & \tabincell{l}{Expected electricity production during a year} & \tabincell{l}{GWh} \\
            \cline{3-5}
			&  & \tabincell{l}{Geometry} & \tabincell{l}{Location of the power plant with latitude and longitude} & \tabincell{l}{NA} \\

            \cline{1-5}
			\multirow{3}{*}{\tabincell{l}{Wind power plants with license under process.geojson}} & \multirow{3}{*}{\tabincell{l}{46}} & \tabincell{l}{Capacity} & \tabincell{l}{Power capacity in design} & \tabincell{l}{MW} \\
			\cline{3-5}
			&  & \tabincell{l}{Name} & \tabincell{l}{Name of the power plant} & \tabincell{l}{NA} \\
            \cline{3-5}
			&  & \tabincell{l}{Expected production} & \tabincell{l}{Expected electricity production during a year} & \tabincell{l}{GWh} \\
			\cline{3-5}
			&  & \tabincell{l}{Geometry} & \tabincell{l}{Location of the power plant with latitude and longitude} & \tabincell{l}{NA} \\

            \cline{1-5}
			\multirow{2}{*}{\tabincell{l}{Wind turbine locations.geojson}} & \multirow{2}{*}{\tabincell{l}{1,458}} & \tabincell{l}{Facility No.} & \tabincell{l}{The index of the power plant that the wind turbine belongs to} & \tabincell{l}{NA} \\
			\cline{3-5}
			&  & \tabincell{l}{Geometry} & \tabincell{l}{Location of the wind turbine with latitude and longitude} & \tabincell{l}{NA} \\

            \cline{1-5}
			\multirow{2}{*}{\tabincell{l}{Wind resource availability.geojson}} & \multirow{2}{*}{\tabincell{l}{196,318}} & \tabincell{l}{Availability} & \tabincell{l}{Availability of wind resource 50 meters above ground during a year} & \tabincell{l}{Hour} \\
			\cline{3-5}
			&  & \tabincell{l}{Geometry} & \tabincell{l}{Polygon of the geographic area of the same availability with latitude and longitude} & \tabincell{l}{NA} \\

            \cline{1-5}
			\multirow{5}{*}{\tabincell{l}{Solar power plants with capacity.geojson}} & \multirow{6}{*}{\tabincell{l}{89}} & \tabincell{l}{Name} & \tabincell{l}{Name of the power plant} & \tabincell{l}{NA} \\
			\cline{3-5}
			&  & \tabincell{l}{Capacity} & \tabincell{l}{Power capacity in design for the power plant} & \tabincell{l}{MW} \\
			\cline{3-5}
			&  & \tabincell{l}{Capacity in operation} & \tabincell{l}{Power capacity of the power plant during operation} & \tabincell{l}{MW} \\
            \cline{3-5}
			&  & \tabincell{l}{Expected production} & \tabincell{l}{Expected power production during a year} & \tabincell{l}{GWh} \\
            \cline{3-5}
            &  & \tabincell{l}{Geometry} & \tabincell{l}{Location of the solar power plant with latitude and longitude} & \tabincell{l}{NA} \\

            \cline{1-5}
			\multirow{3}{*}{\tabincell{l}{Estimated municipality solar power production.geojson}} & \multirow{3}{*}{\tabincell{l}{357}} & \tabincell{l}{Municipality} & \tabincell{l}{Name of municipalities in Norway} & \tabincell{l}{NA} \\
			\cline{3-5}
			&  & \tabincell{l}{Estimation} & \tabincell{l}{Estimated solar power production in a municipality} & \tabincell{l}{MWh} \\
			\cline{3-5}
			&  & \tabincell{l}{Geometry} & \tabincell{l}{Polygon of the geographic boundary of a municipality} & \tabincell{l}{NA} \\

            \cline{1-5}
			\multirow{1}{*}{\tabincell{l}{Solar panel distribution in Oslo.geojson}} & \multirow{1}{*}{\tabincell{l}{868}} & \tabincell{l}{Geometry} & \tabincell{l}{Polygons of rooftop solar panels in Oslo} & \tabincell{l}{NA} \\
            
			\cline{1-5}

		\end{tabular}
	}
\end{table*}

We collected Wind power plant data from Geonorge.no\footnote{The raw data is available at \url{https://kartkatalog.geonorge.no/metadata/vindkraftverk/ac249604-cd82-490c-83cc-9cd24fe18088}, accessed Sep. 19, 2025.} in SHAPE format\footnote{The data is also available in SOS format.}. This data provides wind power plants with licenses or license applications under processing. The data includes the capacities of the power plants, and also provides the locations of wind turbines across Norway, as shown in Fig.~\ref{fig:wind_power_plants}. We converted the data from SHAPE to GeoJSON format using Python in Google Colab, and stored the processed data in Zenodo. We built interactive map based on this data, which is publicly accessible via Online ArcGIS\footnote{The interactive map is available at \url{https://uio-no.maps.arcgis.com/apps/mapviewer/index.html?webmap=065d48122dc843bcbdff2593dc055cc7}, accessed Sep. 19, 2025.}.

We also collected wind power resource data from Geonorge.no\footnote{The raw data is available at \url{https://kartkatalog.geonorge.no/metadata/vindressurser/21079f3d-81b8-405b-bfb1-c213d732fcfb}, accessed Sep. 19, 2025. The data is measured by Kjeller Vindteknikk on behalf of NVE.} in TIF format. We converted the data into GeoJSON format using Python in Google Colab so that it is compatible in GIS platforms. This data presents the availability of wind resources in mainland Norway and the offshore waters, as shown in Fig.~\ref{fig:wind_resource}, with a horizontal resolution of 1x1 km. The data provides the operating hours of turbines in a year at a height of 50 meters above ground. The processed data is stored at Zenodo, and we created an interactive map based on this data, which is accessible via Online ArcGIS\footnote{The interactive map is publicly available at \url{https://uio-no.maps.arcgis.com/apps/mapviewer/index.html?webmap=2321e585a93f46129d196d3ebe7e7a7f}, accessed Sep. 19, 2025.}.



\section{VALIDATION AND QUALITY} 

\begin{figure}[htbp]
\centerline{\includegraphics[width=3.3in]{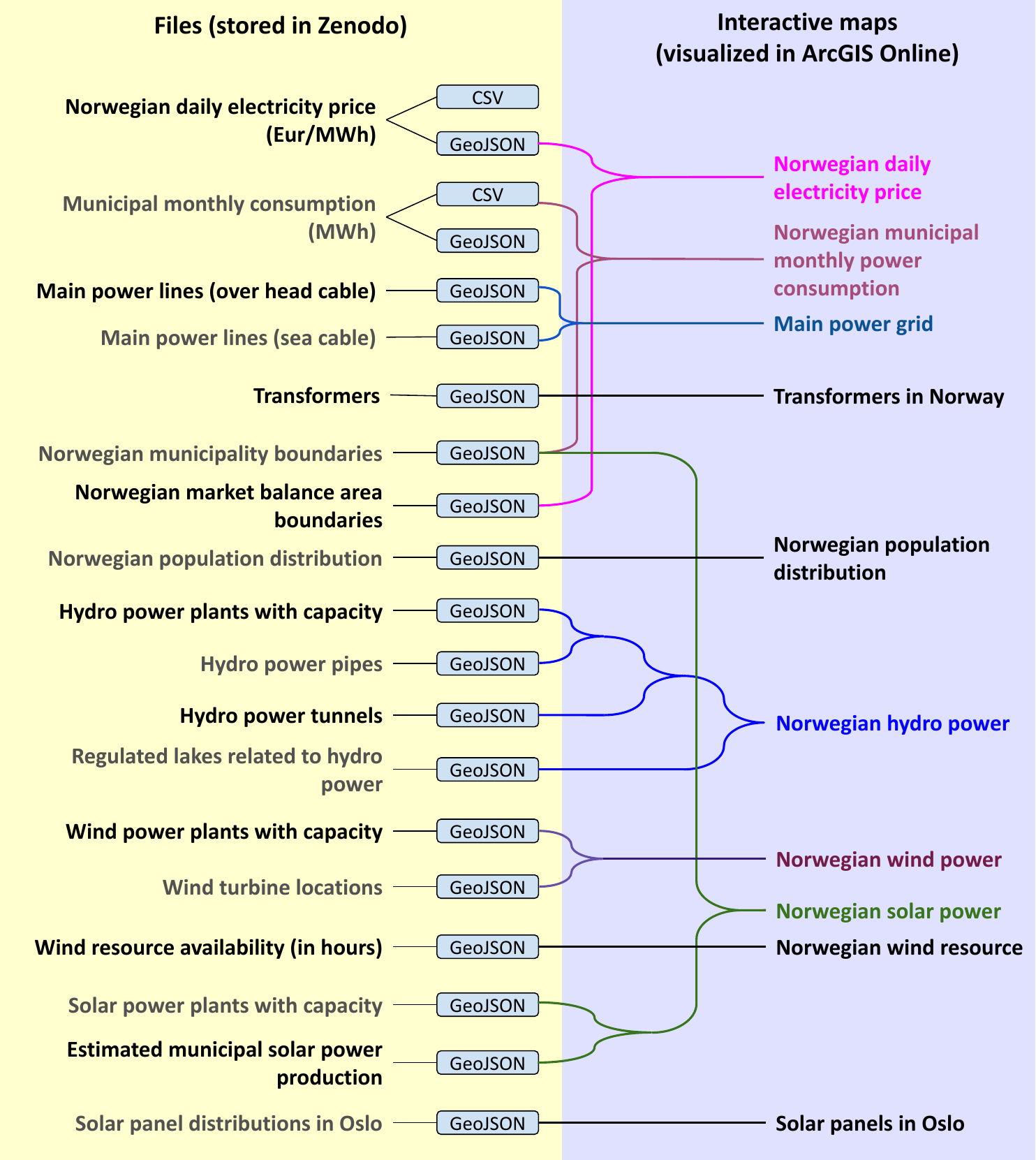}}
\caption{Structure of the files in the generated dataset.\label{fig:structure}}
\end{figure}

We collected most of the raw data from authoritative sources, where the data is either reported/registered by an energy institute (\textit{e.g.}, municipality level energy consumption and  main power grid topology), released by national statistics (\textit{e.g.}, population density), or estimated by governmental institutes (\textit{e.g.}, municipality level solar power production). This provides validity and traceability of our generated dataset. We have also utilized data contributed by individuals, the quality of which we will clarify further. Table~\ref{tbl:data_validity} briefs quality of our data with respect to the data source, creation time, data type, and the accuracy label of the data assigned by us. 

In Table.~\ref{tbl:data_validity}, we view the data as valid and accurate if the data that is real and registered by energy stakeholders or the information that is made transparent to the public by the government. This includes the data of Norwegian daily electricity price, municipality level power consumption, main power grid topology and transformer location and capacity, municipal geographic boundary, geographic electricity price areas, and hydro, wind, and solar power plant data.

We also have data in Table.~\ref{tbl:data_validity} estimated by governmental institutes, \textit{i.e.}, the population density and wind resource availability. These data cannot be provided accurately due to the dynamic nature of the data subject, \textit{e.g.}, population fluidity and intermittence of the wind. It is also impractical to update these data frequently because of the prohibitive large space for data collection. In that, these data is generated in a sampled and statistic sense by governmental institutes. Particularly, the population density data is estimated by SSB based on the registered population linked to address points in the cadastre. The wind resource data has a horizontal resolution of 1x1 km and provides the estimation of wind resource availability (in hours) calculated based on annual mean wind at a height of 50 above ground.

The municipality level solar power production involves power generated from residential rooftops that is not metered, thus the actual production amount is not available. Even though, NVE estimates the solar power production according to registered solar panel installations. The estimation is calculated based on the average weather year in Norway. Therefore, the estimation may deviate from actual municipality level solar power production, and users should leverage this data with caution in their analysis.

Note that in Table.~\ref{tbl:data_validity}, it is impossible to check and validate the solar panel distribution data contributed by individual users in Open Street Map. However, this information is reviewed by a wide range of open-source communities, and the code we released can easily update this data within minutes. This provides a way to gain insights into residential solar power generation that is not measured.

\section{RECORDS AND STORAGE} 

All the generated data is in CSV and/or GeoJSON format. We stored the data with public access in Zenodo \cite{zhang_2025_16794604}\footnote{Our dataset can be accessed at \url{https://doi.org/10.5281/zenodo.16794603}, accessed Sep. 19, 2025.}, and we openly released the code in generating the data in Google Colab.

For each of the dataset file we generated, we provide detailed description of the data, include number of records and the explanation of all the features, as shown in Table.~\ref{tbl:csv_details}. We also illustrate how different data files are related with each other in Fig.~\ref{fig:structure}. From the figure, it is intuitive that we combine different GeoJSON files in creating interactive maps visualized in ArcGIS Online. This combination enriches the information in the interactive maps and can contribute to joint analysis on the power system, \textit{e.g.}, in scheduling energy transmission under constraints of infrastructure, energy availability, local energy demand, electricity price, \textit{etc}.

\section{INSIGHTS AND NOTES} 
From the created data and the interactive maps, it is intuitive that the energy consumption in Norway is closely related to the population distribution. We also observe that solar power generation is significantly higher in the south than the north, while hydro power and wind power are more evenly distributed. Nevertheless, the power grid and transformers connect better and concentrate in the south, while the connection between the south and north can be less robust for power transmission. This might limit the power grid's ability to transfer the surplus energy in the north to the south where the energy demand is higher. Consequently, this infrastructure disparity between the south and north can contribute to the lower energy prices experienced in the north compared to the south.

Beyond the preliminary insights from the dataset created, the dataset is made available in CSV and/or GeoJSON format that can be processed in commonly used programming languages like Python, MATLAB, R, \textit{etc}. We openly release the Python code in generating the data in Google Colab, where the code is readily executable online. The GeoJSON version of the dataset can be recognized in geographic software platforms like ArcGIS and QGIS. The interactive maps are also openly accessible through ArcGIS Online and can be further edited for customized purposes.


\section{SOURCE CODE AND SCRIPTS} 

We developed the code in Phython 3.11 in generating the dataset. The code is publicly available and executable in our Google Colab repository~\cite{code_repository}.


\bibliographystyle{unsrturl}

\bibliography{References}

\end{document}